\documentclass[sigconf, 10pt]{acmart}
\copyrightyear{2025}

\acmConference[WiNTECH '25]{19th ACM Workshop on Wireless Network Testbeds, Experimental evaluation \& Characterization}{November 8, 2025}{Hong Kong}
\acmYear{2025}
\usepackage{graphicx} 
\usepackage{subcaption}
\usepackage{tabularx} 
\usepackage{booktabs} 
\usepackage{soul}
\usepackage{hyperref}
\usepackage{multirow}  
\title{Experimental Insights from OpenAirInterface 5G positioning Testbeds: Challenges and solutions}

\author{Mohsen Ahadi}
\email{mohsen.ahadi@eurecom.fr}
\orcid{0000-0003-1819-5224}
\affiliation{%
  \institution{EURECOM}
  \city{Sophia Antipolis}
  \country{France}
}

\author{Adeel Malik}
\email{adeel.malik@firecell.io}
\orcid{https://orcid.org/0000-0002-7789-9144}
\affiliation{%
  \institution{Firecell}
  \city{Nice}
  \country{France}
}

\author{Omid Esrafilian}
\orcid{https://orcid.org/0000-0002-9720-1815}
\email{omid.esrafilian@eurecom.fr}
\affiliation{%
  \institution{EURECOM}
  \city{Sophia Antipolis}
  \country{France}
}

\author{Florian Kaltenberger}
\orcid{https://orcid.org/0000-0002-3225-3569}
\email{florian.kaltenberger@eurecom.fr}
\affiliation{%
  \institution{EURECOM}
  \city{Sophia Antipolis}
  \country{France}
}

\author{Cedric Thienot}
\email{cedric.thienot@firecell.io}
\affiliation{%
  \institution{Firecell}
  \city{Nice}
  \country{France}
}

\begin{document}

\begin{abstract}
5G New Radio (NR) is a key enabler of accurate positioning in smart cities and smart factories.
This paper presents the experimental results from three 5G positioning testbeds running open-source OpenAirInterface (OAI) gNB and Core Network (CN), using Uplink Time Difference of Arrival (UL-TDoA) with the newly integrated Location Management Function (LMF). The testbeds are deployed across both indoor factories and outdoor scenarios with O-RAN Radio Units (RUs), following a 3GPP-compliant system model. The experiments highlight the impact of synchronization impairments, multipath propagation, and deployment geometry on positioning accuracy. To address these challenges, we propose tailored ToA and TDoA filtering as well as a novel position estimation method based on Particle Swarm Optimization (PSO) within the LMF pipeline. Moreover, we show a beyond-5G framework that leverages non-conventional measurements such as Channel Impulse Response (CIR) to train and test Artificial Intelligence and Machine Learning (AI/ML) models for data-driven positioning.
The results demonstrate the feasibility of achieving 1–2 meter positioning accuracy in 90\% of cases in different testbeds, offering practical insights for the design of robust 5G positioning systems. Moreover, we publicly release the datasets collected in this work to support the research within the 5G positioning community.
\end{abstract}


\maketitle
\section{Introduction}\label{Sec:1}
5G New Radio (NR) marks a major advancement in wireless communications by offering significant improvements in speed, connectivity, and latency. Among its key innovations is high-precision positioning, which enables applications across various sectors—including automotive, logistics, and smart cities \cite{9665436}.
While traditional localization methods such as Global Positioning System (GPS), Bluetooth Low Energy (BLE) and WiFi often face limitations in accuracy and availability particularly in indoor scenarios or dense urban environments, 3GPP offers reliable solutions for positioning in a 5G NR network, by leveraging techniques such as Received Signal Strength Indicator (RSSI), Time of Arrival (ToA), Time Difference of Arrival (TDoA), Angle of Arrival or Departure (AoA/AoD) and Enhance Cell ID (E-CID)\cite{ahadi20235gnr}.
Despite ongoing advancements in enhancing the positioning accuracy through synthetic data in simulation environments, fewer contributions focus on validating these methods in a standardized and reproducible manner using real-world testbeds. This is an essential step for uncovering the practical challenges that arise during deployment.

In our previous work \cite{malik2025conceptreality5gpositioning}, we presented an open-source implementation of the New Radio Positioning Protocol A (NRPPa)\cite{3gpp_ts38_455} along with a Location Management Function (LMF)\cite{3gpp_ts29_572} running in OpenAirInterface (OAI). OAI provides an open-source implementation of the 5G protocol stack for both Radio Access Network (RAN) and Core Network (CN) components \cite{Kaltenberger2024}. We validated the end-to-end procedure for Up Link TDoA positioning, using an O-RAN–compliant testbed, including a mobile phone serving as User Equipment (UE) and commercial O-RAN Radio Units (O-RUs).

In this paper, we present a complete positioning pipeline implemented on the OAI LMF, which includes filtering of ToA and TDoA measurements using empirical and geometric criteria, and a Particle Swarm Optimization (PSO) based position estimation algorithm. We evaluate its performance on three diverse testbeds: the outdoor GEO-5G testbed at EURECOM, and two indoor industrial sites at STELLANTIS' Mattern Lab and the Airbus factory hall. These environments pose distinct challenges, such as synchronization impairments in outdoor setups with widely distributed RUs, and severe multipath and NLoS conditions indoors.
Our PSO-based estimator is designed to handle timing offsets in multi-RU deployments, while the ToA and TDoA filters help to discard unreliable measurements. We also introduce a beyond-3GPP framework for evaluating a data-driven positioning within the O-RAN framework, using Channel Impulse Response (CIR) as input to improve robustness in challenging scenarios. Results demonstrate 1–2 meter accuracy in 90\% of cases across all testbeds. An extensive dataset comprising measured  CIRs, precise timestamps, and corresponding position labels is published alongside this work.
\section{Related Works}
A range of studies have investigated positioning capabilities on the OpenAirInterface (OAI) platform, utilizing diverse techniques and setups. For instance, earlier works have presented tools and initial frameworks to support positioning in OAI \cite{10925986,10757829}, explored E-CID mechanisms \cite{9880817}, and proposed hybrid GNSS and 5G NR downlink PRS-based positioning for both UAV and terrestrial gNBs \cite{ENC2023-15432}. Other research has focused on enhancing Round-Trip Time (RTT) estimation accuracy \cite{gangula2024roundtriptimeestimation}, implementing fingerprinting approaches based on RSRP measurements \cite{8845272}, and evaluating sidelink-based localization strategies \cite{10.1145/3696380}. More experimental studies include a single-link testbed emulating a multi-node TDoA setup using software-defined radios (SDRs) \cite{palama2024experimental}.
Despite these contributions, to the best of our knowledge, our work is the first to report positioning results from a fully open-source, 3GPP-compliant implementation on OAI, including both OAI RAN and CN components, including LMF. Furthermore, our framework has been validated on commercial RUs in various environments, demonstrating its readiness for practical deployment. This paper presents the signal processing methods and experimental outcomes from these deployments.
\section{OAI Positioning System Model}
We deploy 5G Stand Alone (SA) networks using OAI gNBs and CN, following a disaggregated O-RAN architecture (split option 7.2) with Central Unit (CU), Distributed Unit (DU), and multiple RUs per gNB. Each RU is equipped with multiple fixed antennas and synchronized via Precision Timing Protocol (PTP) using a Grandmaster clock over the Global Navigation Satellite System (GNSS), achieving nanosecond-level timing precision.
A UE with fixed height moves in $t \in \{1,T \}$ timestamps with unknown position \(\mathbf{u}_t \in \mathbb{R}^3\), always bounded within the convex frame of $M$ antennas each having known location $\mathbf{x_m}\in\mathbb{R}^3$. At each timestamp, the UE transmits UL Sounding Reference Signal (SRS), received by antennas.
The system operates under Orthogonal frequency-division multiplexing (OFDM) with \(N_{\text{fft}}\) subcarriers. The CIR \(\mathbf{h}_{m,t}\) is computed via IDFT from the estimated Channel Frequency Response (CFR) vector \(\mathbf{w}_{m,t}\). The ToA of the signal at antenna $m$ and time index $t$ is estimated as the delay corresponding to the strongest multipath component, identified by the peak magnitude of the CIR vector. Specifically, the ToA in seconds is given by:
\begin{equation}
\tau_{m,t} = T_s \cdot \arg\max_n \left| \mathbf{h}_{m,t}[n] \right|,
\end{equation}
where \(T_s\) is the sampling period in second/sample unit.
The set of all ToAs and antenna locations are vectorized as \(\boldsymbol{\tau}_t \in \mathbb{R}^M\) and \(\mathbf{X} \in \mathbb{R}^{M \times 3}\) respectively, and retrieved by the LMF over the NRPPa positioning procedure. To eliminate the UE-side clock offset, the LMF calculates TDoA by differencing ToAs with respect to a reference antenna. These TDoA measurements, along with the antenna coordinates, are used to estimate the UE’s position. Details on the TDoA-based estimation framework are presented in Section~\ref{sec:tdoa_positioning}.
\begin{figure}
    \centering
    \includegraphics[width=0.99\linewidth]{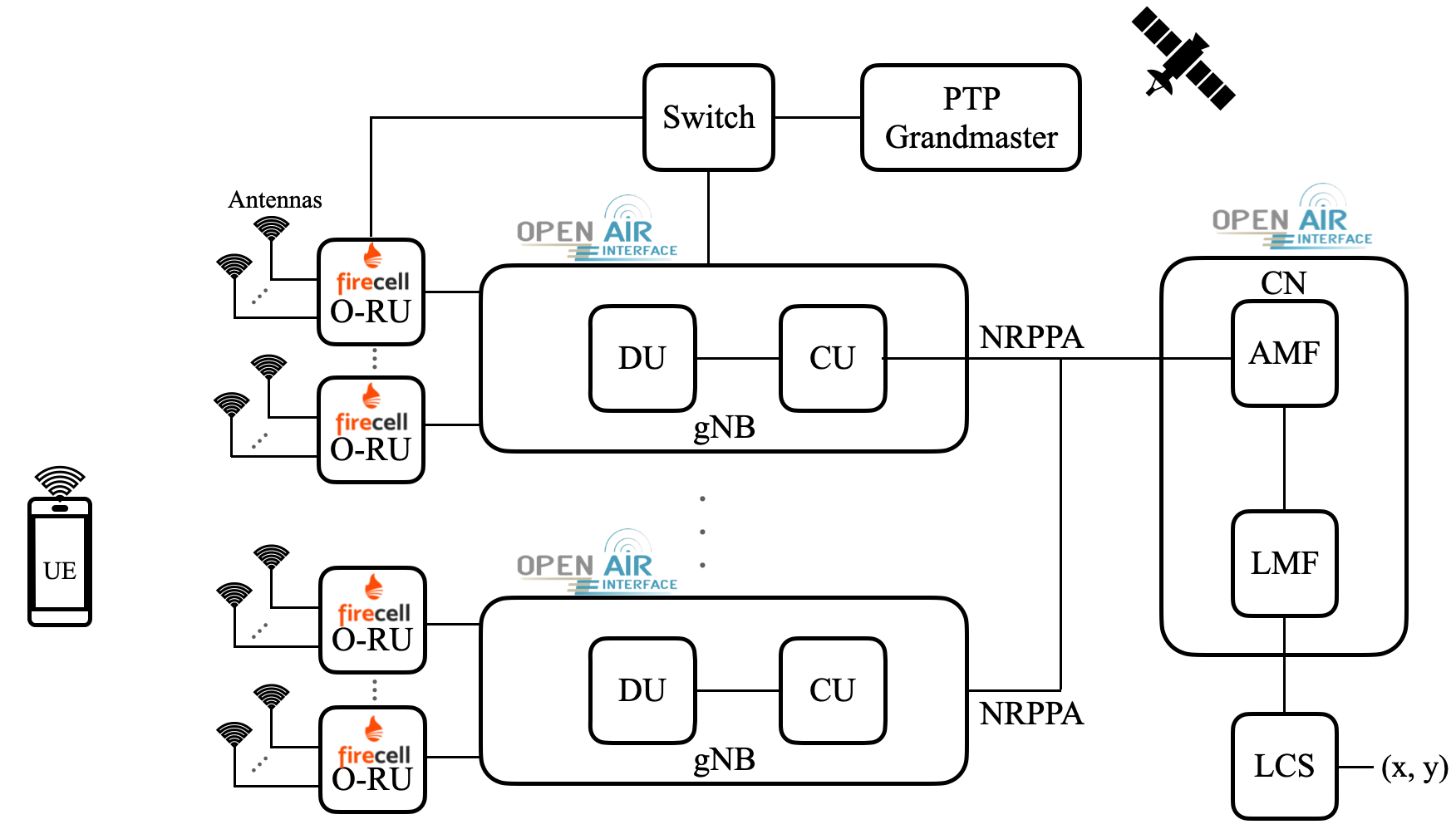}
    \caption{System model with OAI CN, multiple OAI gNBs, multiple RUs, and distributed antennas for UL-SRS TDoA positioning}
    \Description{}
    \label{fig:systemmodel_tdoa}
\end{figure}

\section{Our 5G Positioning Testbeds}
Building on the aforementioned principles of the UL-TDoA positioning system model, this section presents three distinct localization testbeds.

\subsection{GEO5G Outdoor Deployment at EURECOM}
\begin{figure}[t]
    \centering
    \begin{minipage}[t]{0.4\textwidth}
        \centering
        \includegraphics[width=\textwidth]{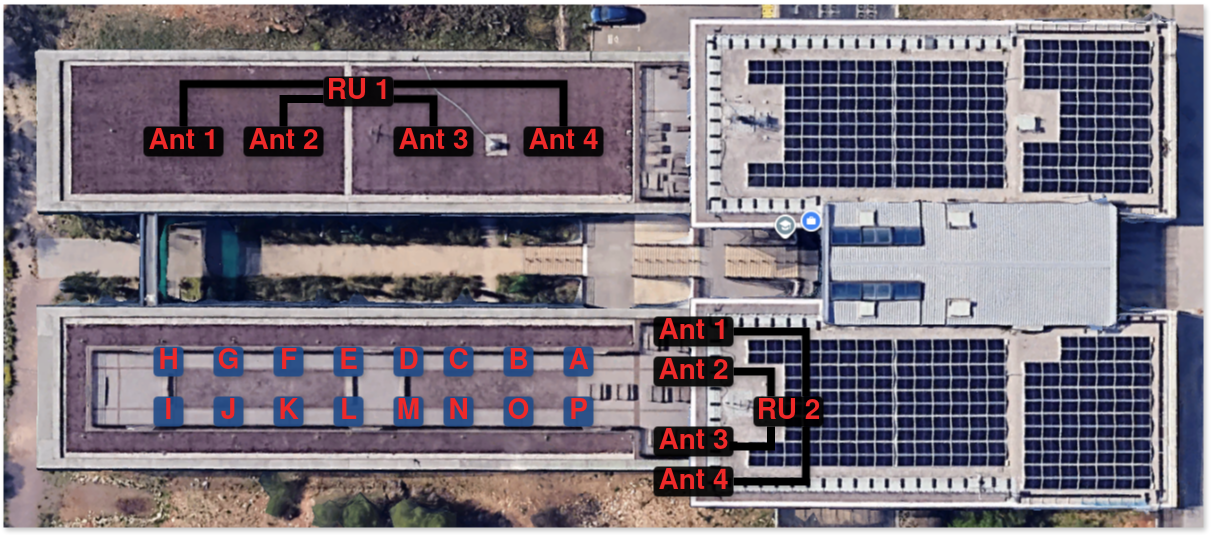}
        \caption*{(a) GEO-5G testbed at EURECOM}
        \label{fig:eurecom_map}
    \end{minipage}%
    \vspace{2mm}
    \begin{minipage}[t]{0.2\textwidth}
        \includegraphics[width=\textwidth]{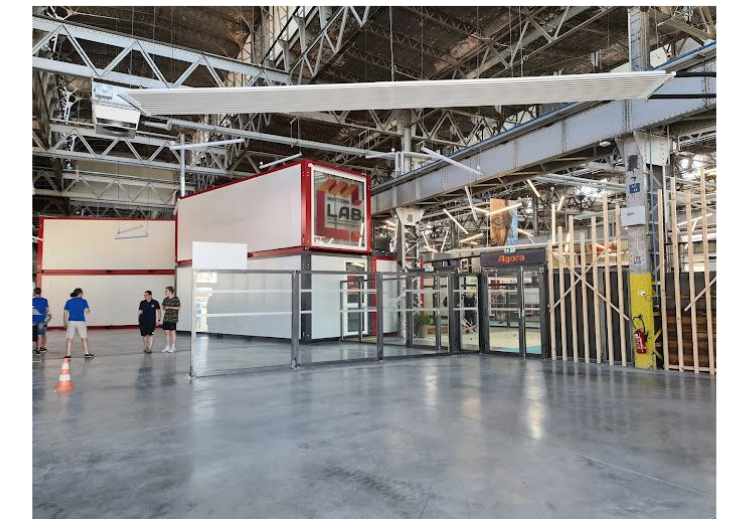}
        \caption*{(b) Stellantis testbed}
        \label{fig:stellantis_map}
    \end{minipage}%
    \begin{minipage}[t]{0.2\textwidth}
        \includegraphics[width=\textwidth]{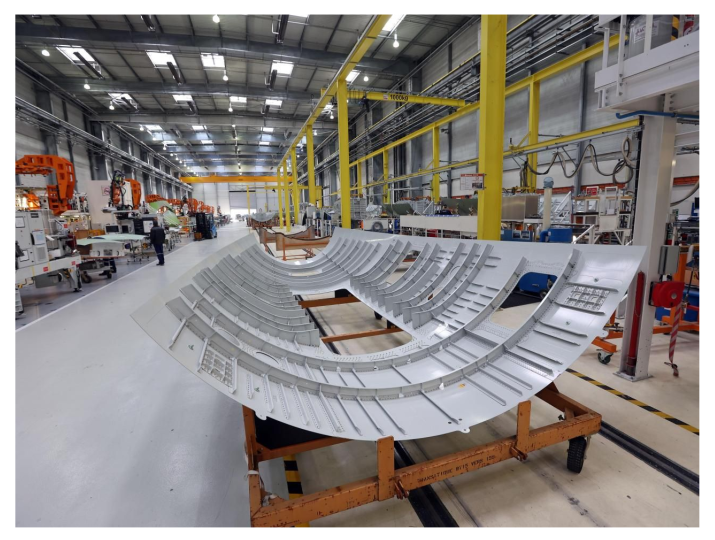}
        \caption*{(c) Airbus factory}
        \label{fig:airbus_map}
    \end{minipage}
    \caption{Overview of 5G positioning testbeds}
    \label{fig:testbeds}
    \Description{}
\end{figure}

The GEO5G testbed is part of EURECOM’s Open5G platform, featuring high-speed fiber infrastructure and supporting virtualized 5G with USRPs~\cite{ettus_usrp}, O-RAN RUs, and OAI. To evaluate OAI’s new localization features, we integrated two outdoor O-RAN RUs provided by Firecell~\cite{firecell_5g} from VVDN~\cite{vvdn_oran_portfolio_2023} into the testbed. Our contributions in ~\cite{malik2025conceptreality5gpositioning} allow flexible deployment using either single-gNB or multi-gNB configurations with multiple RUs.
Antennas are distributed across RUs and mounted on roof railings using low-loss cables, covering a 50\,m$\times$10\,m area on EURECOM’s north terrace (see Fig.~\ref{fig:testbeds}a). Ground truth positions at this testbed are determined using two methods. For static positioning scenarios, where the UE is mounted on a tripod to evaluate optimal performance, 16 test points from A to P in Fig.\ref{fig:testbeds}a are distributed across the testbed. Their precise coordinates are measured using laser equipment by recording distances from each test point to all antennas and computing the positions through geometric triangulation.
For a handheld mobile UE that takes a trajectory freely in the testing area, we set up a Real-Time Kinematic (RTK) setup that includes a base GPS module that is steady and a rover GPS module that moves closely with the UE. RTK positioning enables centimeter-level accuracy by using carrier-phase measurements and differential corrections from its base~\cite{misra2011gps}.
In order to convert RTK geographic coordinates from [lat, long, alt] to a raw Cartesian coordinates [xEst, yNorth, zUp], we use open-source libraries such as \textit{pymap3d} in real-time on Python. A linear transformation is then applied to align these with our local [x, y, z] coordinate system, with antenna 1 on RU 1 as reference [0,0,2.2].
The testbed follows an O-RAN architecture with synchronization option LLS-C3, using a GNSS-disciplined Qulsar Qg~2 Grandmaster. The CU/DU operates on a high-performance server connected via Cisco switches, while the CN runs in Docker on a separate server.
Precise synchronization is critical for TDoA, as 1\, ns error translates to 0.3\,m inaccuracy. However, due to PTP switch impairments and RU long distances at this testbed, errors up to 40\, ns are observed. To mitigate the propagation of clock drift from one RU to the other, we adopt a per-RU reference antenna strategy in calculating TDoAs instead of a common reference across all RUs, ensuring internal consistency and reducing cross-RU time drift. This modification alters the classical TDoA-based positioning, detailed in Section~\ref{sec:tdoa_positioning}.
\subsection{STELLANTIS Indoor Deployment}
\begin{figure}[t]
    \centering
    \begin{subfigure}[b]{0.20\textwidth}
        \includegraphics[width=\textwidth, height=\textwidth]{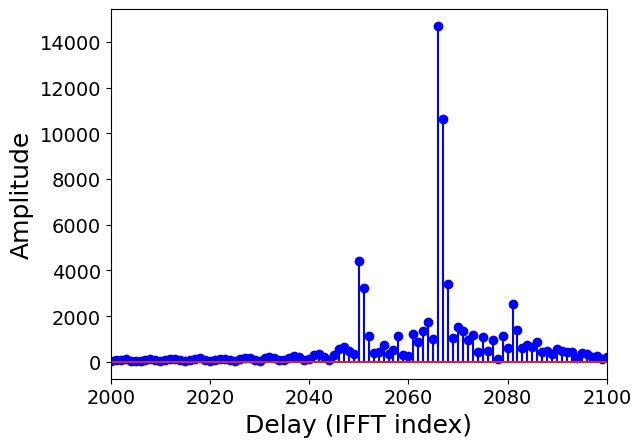}
        \caption{Outdoor CIR example from EURECOM}
        \label{fig:outdoor}
    \end{subfigure}
        \begin{subfigure}[b]{0.20\textwidth}
        \includegraphics[width=\textwidth, height=\textwidth]{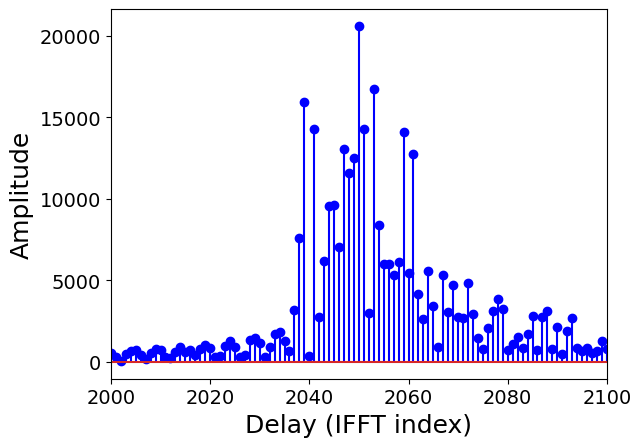}
        \caption{Indoor CIR example from Mattern lab}
        \label{fig:indoor}
    \end{subfigure}
    \caption{Outdoor/indoor multipath  CIR}
    \label{fig:outdoorindoorcir}
        \Description{}
\end{figure}

A similar deployment to EURECOM is installed at STELLANTIS, located in the Mattern lab (Sochaux, France), depicted in Fig.\ref{fig:testbeds}b, with a network architecture similar to Fig.~\ref{fig:systemmodel_tdoa}. This indoor setup includes 2 gNBs, 4 RUs and 8 antennas, covering a \(20\,\text{m} \times 20\,\text{m}\) area. While fewer nodes would cover the area, 2 gNBs and 4 RUs were deployed to test both single and multi-gNB configurations at scale.
Despite LoS connectivity to all antennas, the small environment and indoor propagation pose challenges for positioning. Multipath components exhibit minimal delay separation, causing closely spaced peaks in the CIR—unlike the more separable multipath in outdoor environments (see Fig.~\ref{fig:outdoorindoorcir}).
In literature, super-resolution methods such as MUSIC~\cite{1259415} have been studied for ToA estimation in such dense environments, but their reliance on prior multipath knowledge and high computational cost limit their suitability for real-time tracking applications.
In contrast, we propose a low-complexity, low-latency pre-processing step, detailed in Section\ref{sec:tdoa_positioning}. While the legacy OAI channel estimation applies interpolation and oversampling before ToA estimation—helping to detect peaks in closely spaced multipath scenarios—we introduce additional ToA and TDoA filtering mechanisms at the LMF. These filters improve positioning robustness without incurring the computational overhead of super-resolution algorithms.
\subsection{Airbus factory hall deployment}
Ultimately, the Airbus Atlantic factory hall in Méaulte, France, is deployed with a similar system model in Fig. \ref{fig:systemmodel_tdoa} and equipped with $2$ gNBs, $4$ RUs, and $16$ antennas. Fig.\ref{fig:testbeds}c has a sample image of the Airbus factory hall where the antennas are covering an area of 100m$\times$50m. However, this deployment faces significant signal quality challenges. The long 50 m cables connecting the antennas to the RUs introduce considerable Signal-to-Noise Ratio (SNR) degradation. Additionally, the presence of large metallic structures and sizable airplane frames under construction further complicates the propagation environment. These obstructions severely limit the availability of LoS conditions, despite the high number of deployed antennas. As a result, the raw received signals and channel estimations are heavily affected by noise and multipath. Ensuring reliable positioning performance in such environments requires specialized measurement filtering techniques. In section \ref{sec:tdoa_positioning}, we introduce a set of filters designed based on the known positions of the antennas and the geometry of the testing area, as well as the delay spread and of favorable signals to suppress noisy or NLoS measurements. While the proposed processing pipeline yields accurate results at some testing points with enough LoS antennas, the TDoA-based method at the Airbus testbed exhibits high localization errors across the majority of the testing area.

To address this limitation, we propose a data-driven method based on Finger Printing, where in a large dataset, each set of CIRs from all antennas is labeled with its corresponding ground truth position calculated by a laser tool. Although this approach is inherently exhaustive and requires significant effort for data labeling and a large volume of training samples, it demonstrates improved positioning reliability compared to the traditional TDoA-based method. By leveraging the power distribution knowledge in LoS/NLoS CIRs, the Finger Printing method, enhanced by the feature extraction capabilities of Convolutional Neural Networks (CNNs), demonstrates greater robustness in complex environments, such as the Airbus testbed. The complete framework and the necessary pre-processing steps enabling the Finger Printing approach are detailed in Section~\ref{sec:FP}
\section{Position estimation with refined ToA and TDoA}\label{sec:tdoa_positioning}
\subsection{ToA Filtering}
\begin{figure}
    \centering
    \includegraphics[width=0.85\linewidth]{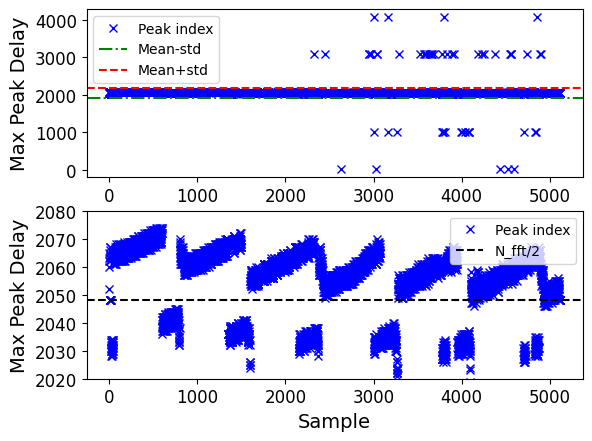}
    \caption{Filtering ToAs by statistical bounds based on max peak delay distribution concentrated around $\frac{N_{\text{fft}}}{2}$}
    \label{fig:peakdelay_dist}
        \Description{}
\end{figure}

Although absolute ToA is not available in our testbeds due to the lack of synchronization between the UE and RUs, an empirical distribution of the channel's delay spread can still be obtained by analyzing large CIR datasets. By applying an IDFT to the FFT-shifted CFR, we process CIRs so that dominant peaks typically align around the center index \( \frac{N_{\text{fft}}}{2} \). However, we observe that some CIRs exhibit peaks with large offsets, leading to physically implausible TDoA values given the known geometry of the test environment. This is mainly attributed to synchronization impairments that introduce timing jitter.
Fig.~\ref{fig:peakdelay_dist} shows the distribution of maximum peak locations in the FFT-shifted CIR dataset. The majority of the peaks are concentrated around $\frac{N_{\text{fft}}}{2}=2048$ while a noticeable portion exhibits significant offsets that indicate outliers.

Therefore, we design a ToA filter that preserves CIRs with peaks falling within the expected delay spread range and discards those with excessive temporal offsets.
CIR is modeled as a sum of multipath components, each represented by a delayed Dirac delta function $\delta(.)$ scaled by a complex amplitude $\alpha_i$, which captures both the attenuation and phase shift of the $i$-th path.
\begin{equation}
    h(t) = \sum_{i=1}^{N_{\text{fft}}} \alpha_i \, \delta(n - \tau_i)
\end{equation}
The delays $\tau_i$ correspond to the ToA of each path. 
To formalize the filtering process, let \(\mu_\tau\) and \(\sigma_\tau\) denote the empirical mean and standard deviation of the max-peak delay distribution, respectively. A CIR is retained if its most significant peak \(\tau_{\text{peak}}\) satisfies the condition:
\begin{equation}
\mu_\tau - \sigma_\tau \leq \tau_{\text{peak}} \leq \mu_\tau + \sigma_\tau.
\end{equation}
This range captures typical delay spreads while excluding outliers inconsistent with the environment's propagation.

\subsection{TDoA Filtering}

The geometry between each antenna pair and their direct path to the UE determines whether the TDoA is positive, negative, or zero, depending on the relative positions of the UE and antennas. However, Theorem\ref{thm:tdoa_bound} implies that the absolute value of TDoA is bounded regardless of sign.
\begin{theorem}
\label{thm:tdoa_bound}
Let points \( A \) and \( B \) be two antennas and \( C \) be a UE, such that the three points \( A, B, C \) form a non-degenerate triangle. Denote the distances between the points as \( AB = c \), \( AC = b \), and \( BC = a \).
Then, the absolute difference \( |a - b| \) is bounded as:
\begin{equation}
    0 \leq |a - b| \leq c,
\end{equation}
where the lower bound is trivial due to the non-negativity of absolute values, and the upper bound follows from the triangle inequality in Euclidean geometry \cite{khamsi2001metric}.
Equality in the upper bound holds if and only if the points \( A, B, C \) are collinear (i.e., the triangle is degenerate).
\end{theorem}
In our setup with \(K\) RUs, each with \(M_k\) distributed antennas, we define \(\Delta \tau_{k,m}^{t}\) as the TDoA at time \(t\) for the \(m\)-th antenna of the \(k\)-th RU, relative to a reference antenna within the same RU:
\begin{align}
\Delta{\tau}_{k,m}^t &= \tau_{k,m}^t - \tau_{k,\mathrm{ref}}^t, \label{eq:tdoa} \\
&\quad k \in \{1, \dots, K\},
 m \in \{1, \dots, M_k\}, m \neq \mathrm{ref} \notag.
\end{align}
Here, \(\tau_{k,m}\) denotes the ToA at the \(m\)-th antenna $\mathbf{x}_{k,m}$, and \(\tau_{k,\mathrm{ref}}\) represents the ToA at the reference antenna $\mathbf{x}_{k,\text{ref}}$, both associated with the \(k\)-th RU.
Based on the assumption in our system model, the UE's position is spatially bounded by the convex region of the antennas. 
In the edge cases where the UE is co-located with one of the antennas, the TDoA becomes

\begin{equation}
\Delta \tau_{k,m}^t(\mathbf{u}_t) =
\begin{cases}
    - \dfrac{\| \mathbf{x}_{k,m} - \mathbf{x}_{k,\mathrm{ref}} \|_2}{v}, & \text{if } \mathbf{u}_t = \mathbf{x}_{k,m}, \\
    \phantom{-} \dfrac{\| \mathbf{x}_{k,m} - \mathbf{x}_{k,\mathrm{ref}} \|_2}{v}, & \text{if } \mathbf{u}_t = \mathbf{x}_{k,\mathrm{ref}}.
\end{cases}
\end{equation}
Therefore, we can write
\begin{equation}
0 \leq |\Delta{\tau}_{k,m}|
\leq \frac{\| \mathbf{x}_{k,m} - \mathbf{x}_{k,\mathrm{ref}} \|_2}{c}.
\label{eq:tdoa_bounds}
\end{equation}
Based on the theoretical bounds established in \eqref{eq:tdoa_bounds}, from Theorem~\ref{thm:tdoa_bound}, 
we define a filtering criterion that systematically discards invalid TDoA values. In particular, this filter is designed to eliminate measurements that fall outside the physically feasible range, which typically arise from noisy conditions or NLoS propagation. By enforcing this constraint, we ensure that only reliable TDoA measurements are retained and passed to the positioning algorithm, thereby improving its robustness and accuracy.
\begin{figure}[t]
    \centering
    \includegraphics[width=0.99\linewidth]{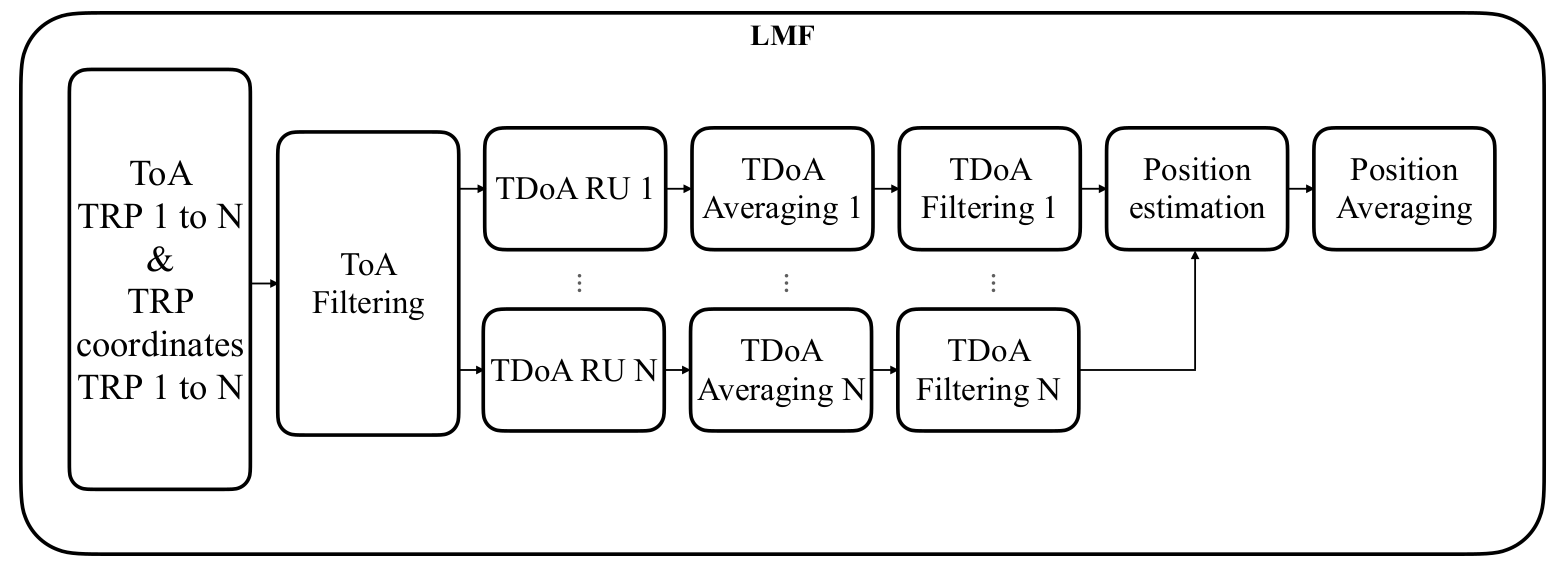}
    \caption{Our proposed positioning pipeline at LMF}
    \label{fig:diagram_positioning}
        \Description{}
\end{figure}
\subsection{Position Estimation}
\subsubsection{Standard TDoA method on LMF}
To estimate the UE position \(\hat{\mathbf{u}} = (\hat{x}, \hat{y})\), we define a TDoA-based loss function on the LMF by comparing the measured and expected TDoAs using known antenna coordinates. For a candidate position \(\hat{\mathbf{u}}_t\), the loss function is defined as
\begin{equation}
\mathcal{L}(\hat{\mathbf{u}}_t) = \sum_{k=1}^{K} \sum_{m=1}^{M_k} \left( {\Delta \hat\tau}_{k,m}^n - \Delta{\tau}_{k,m}(\hat{\mathbf{u}_t}) \right)^2,
\end{equation}
where \(\Delta{\hat\tau}_{k,m}^n\) is the measured TDoA and
\begin{equation}
\Delta{\tau}_{k,m}(\hat{\mathbf{u}}_t) = \frac{\|\hat{\mathbf{u}}_t - \mathbf{x}_{k,m}\| - \|\hat{\mathbf{u}}_t - \mathbf{x}_{k,\text{ref}}\|}{v}
\end{equation}
is the expected TDoA.
To minimize this loss, based on the comparisons to state-of-the-art methods in our recent work\cite{ahadi20235g}, we employ PSO due to its efficiency and lack of dependence on gradient information, which makes it suitable for real-time applications on resource-constrained LMFs. Unlike gradient search or Least Squares methods in \cite{1201741} and \cite{301830}, PSO offers a favorable trade-off between latency and accuracy.
Each particle $p\in\{1, P\}$ represents a candidate UE position and is constrained to move within the bounding box defined by the minimum and maximum antenna coordinates or testing area. At each iteration $i$, the particles' velocity and position are updated as
\begin{align}
\mathbf{v}_p^{(i+1)} &= w \mathbf{v}_p^{(i)} + c_1 r_1 (\mathbf{u}_p^{\text{local}} - \mathbf{u}_p^{(i)}) + c_2 r_2 (\mathbf{u}^{\text{global}} - \mathbf{u}_p^{(i)}), \\
\mathbf{u}_p^{(i+1)} &= \mathbf{u}_p^{(t)} + \mathbf{v}_p^{(i+1)},
\end{align}
where $\mathbf{u}_p^{(i)}$ is the position and \(\mathbf{v}_p^{(i)}\) is the velocity of each particle, \(w = 0.9\) is the inertia weight, \(c_1 = 0.5\), \(c_2 = 0.9\) are acceleration coefficients, and \(r_1, r_2 \sim U(0,1)\) are random variables. \(\mathbf{u}_p^{\text{local}}\) is the best position found by particle \(p\), and \(\mathbf{u}^{\text{global}}\) is the best position found by the swarm.
The estimated UE position is given by \(\mathbf{u}^{\text{global}}\) at the final iteration. In addition, a moving average filter is applied to the output of the PSO to smooth the UE location estimate. The averaging window size plays a significant role in the latency of the position estimations, as larger windows make the response slower but result in smoother trajectories, while smaller windows reduce latency at the cost of increased noise.
By incorporating prior knowledge of the deployment geometry, filtering noisy measurements, and explicitly accounting for hardware impairments in the loss function, the proposed PSO-based method demonstrates robust and reliable performance in real-world positioning testbeds.
In Section \ref{sec:results}, we present the positioning results of the proposed processing pipeline across our various testbed environments.
\begin{figure}[t]
    \centering
    \begin{subfigure}[b]{0.2\textwidth}
        \includegraphics[width=\textwidth, height=\textwidth]{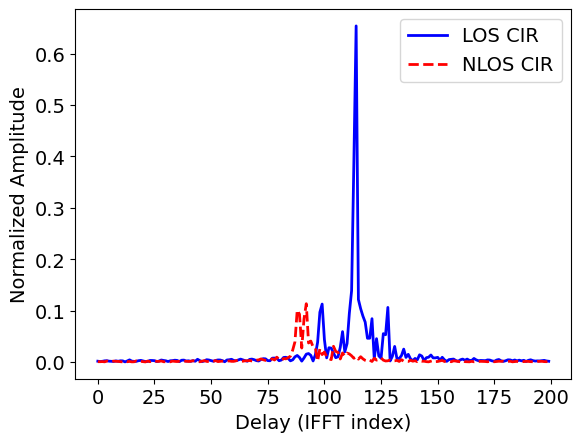}
        \caption{LoS/NLoS sample CIR from testbeds}
        \label{fig:outdoorCIR}
    \end{subfigure}
        \begin{subfigure}[b]{0.2\textwidth}
        \includegraphics[width=\textwidth, height=\textwidth]{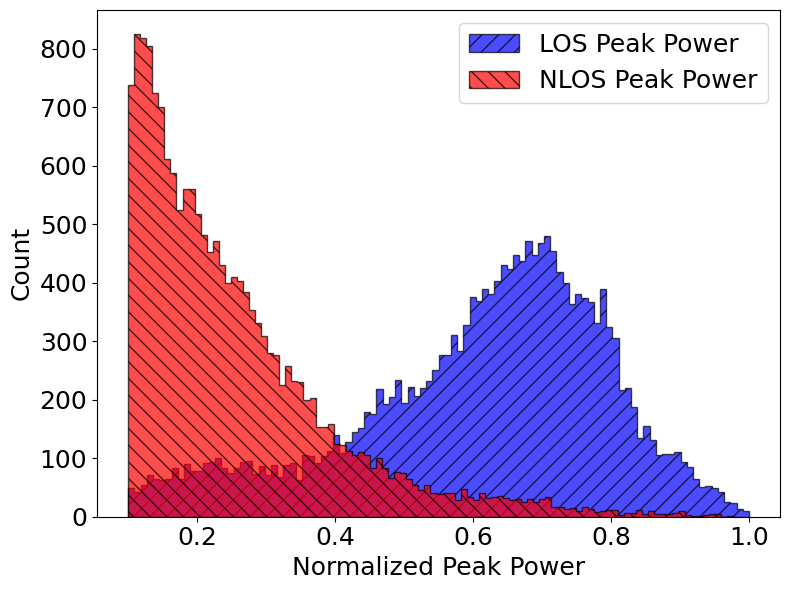}
        \caption{LoS/NLoS max peak power histogram}
        \label{fig:indoorCIR}
    \end{subfigure}
    \caption{LoS/NLoS power distributions}
    \label{fig:losnlos_cir_hist}
        \Description{}
\end{figure}
\subsubsection{Data-driven method by Finger Printing}\label{sec:FP}
While TDoA is widely used for positioning, its accuracy drops in NLoS conditions. In contrast, ML-based methods leveraging richer inputs like full CIR have shown better performance in such environments. Although 3GPP NRPPa \cite{3gpp_ts38_455} does not support large-scale CIR transmission, the O-RAN architecture allows forwarding CIR from the CU-DU to the Near-RT RIC via the E2 interface and xApps, enabling the deployment of learning-based positioning approaches.

We present a positioning procedure inspired by the O-RAN framework, where CIR data is offloaded from gNBs to an external host for AI/ML-based inference. Instead of using E2, our implementation relies on the MQTT protocol. Here, each gNB publishes a topic for CIR measurements to a MQTT broker. The broker can reside on a gNB, the CN, or a third server. An AI/ML host subscribes to this topic and collects CIR data in real time. A pretrained Finger Printing model is deployed on this host to infer the UE’s position. To enable effective inference, raw CIR is first preprocessed into a format suitable for the Finger Printing model.

\begin{figure}
    \centering
    \includegraphics[width=0.99\linewidth]{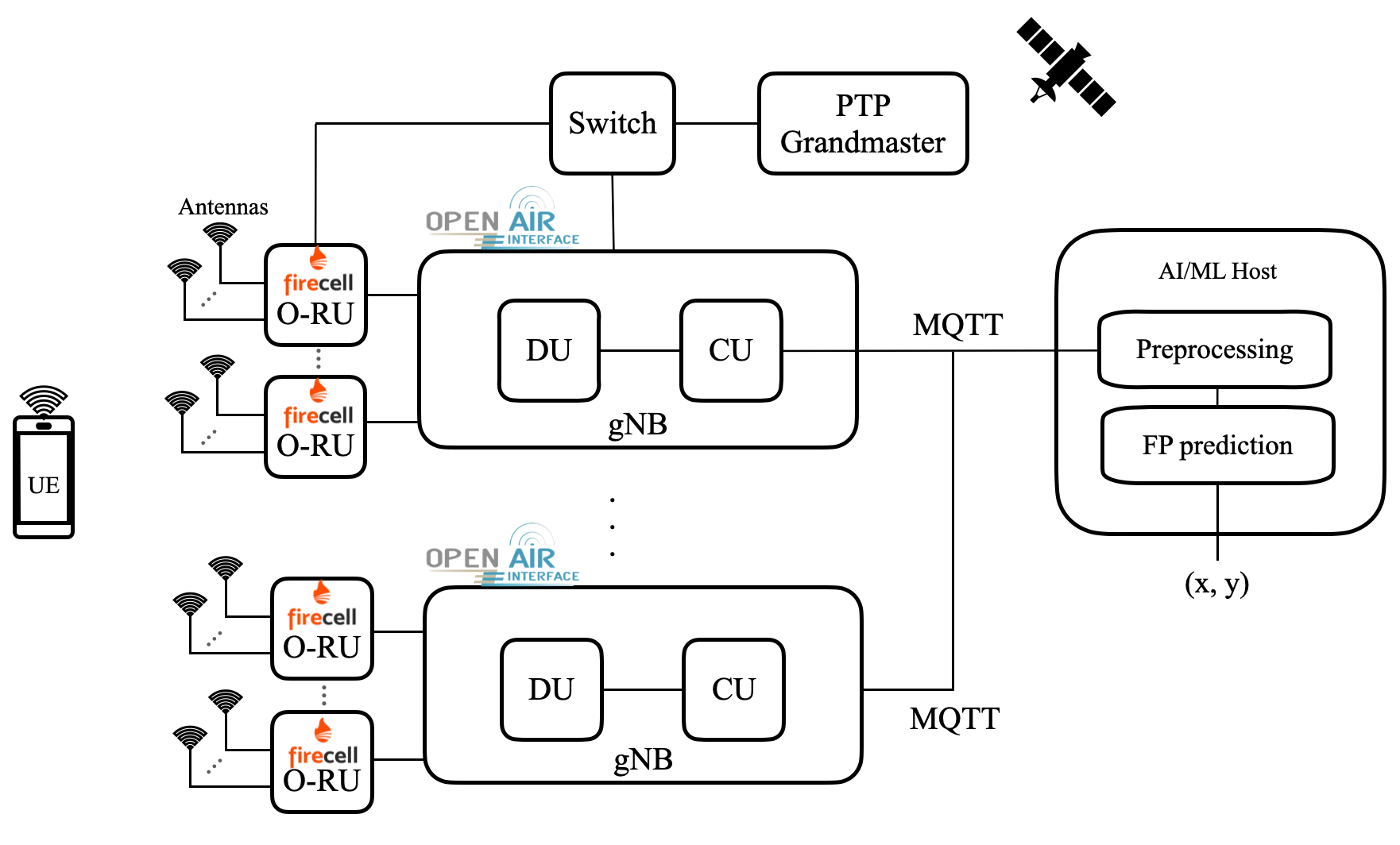}
    \caption{System model for AI/ML positioning}
    \label{fig:fp_diagram}
        \Description{}
\end{figure}

Let the matrix \(\mathbf{H}_{k,t} \in \mathbb{C}^{M_k \times N_{\text{fft}}}\) be the collection of CIRs received from all \(M_k\) distributed antennas of RU \(k\) at timestamp \(t\), where each row corresponds to a single antenna's element-wise absolute value of CIR
\begin{equation}
\mathbf{H}_{k,t} =
\begin{bmatrix}
|\mathbf{h}_{k,1,t}^T| \\
|\mathbf{h}_{k,2,t}^T| \\
\vdots \\
|\mathbf{h}_{k,M_k,t}^T|
\end{bmatrix}
\in \mathbb{C}^{M_k \times N_{\text{fft}}}.
\end{equation}
On each RU, we compensate for the TDoA between antennas by introducing a peak-based alignment step.
Let \(\tau_{k,m,t}^{\text{peak}}\) denote the index of the maximum peak in the \(m\)-th CIR vector \(\mathbf{h}_{k,m,t}\), and an offset index as the earliest peak among all antennas, $\eta_{k,t}^{\text{offset}} = \min_{m \in \{1, \dots, M_k\}} \tau_{k,m,t}^{\text{peak}}$.
This operation shifts the vector to the left and fills the tail with zeros
\begin{equation}
    \mathbf{h}_{k,m,t}^{\text{shifted}}[j] = 
\begin{cases}
\mathbf{h}_{k,m,t}[j + \eta_{k,t}^{\text{offset}}] & \text{if } j + \eta_{k,t}^{\text{offset}} < N_{\text{fft}}, \\
0 & \text{otherwise},
\end{cases}
\end{equation}
resulting in a TDoA-aligned CIR matrix \({\mathbf{H}}_{k,t}^{\text{shifted}} \in \mathbb{C}^{M_k \times N_{\text{fft}}}\).
This procedure, during both training and testing phases, ensures that the CIRs are temporally aligned across antennas of the same RU, thereby respecting the assumption of tight time synchronization among them and the correctness of TDoA.
The matrix \({\mathbf{h}}_{k,t}^{\text{shifted}}\) is further normalized by the maximum peak magnitude observed across the entire training dataset.
To form a unified input dimension for model training and testing, CIRs are concatenated from all \(K\) RUs along the antenna (row) dimension and truncated to only contain the first $C$ fft indices 
\begin{equation}
\mathbf{H}_{t}^{\text{norm}} = \frac{1}{\alpha_{\text{norm}}}.
\begin{bmatrix}
\mathbf{h}_{1,t}^{\text{shifted} \hspace{1mm}T} \\
\mathbf{h}_{2,t}^{\text{shifted} \hspace{1mm}T} \\
\vdots \\
\mathbf{h}_{M,t}^{\text{shifted} \hspace{1mm}T}
\end{bmatrix}
\in \mathbb{C}^{M \times C},
\end{equation}
where the normalization factor \(\alpha_{\text{norm}} = \max\limits_{k,m,t} ({\mathbf{h}}_{k,m,t}^{\text{shifted}})\) is computed from the training data and reused during testing to ensure consistency.
Also, \(M = \sum_{k=1}^K M_k\) is the total number of antennas across all RUs. 
To mitigate the impact of NLoS antennas, a binary masking vector \(\mathbf{m} \in \{0,1\}^{M \times 1}\) is applied to the normalized CIR input matrix. The masking vector suppresses unreliable antennas by zeroing out the corresponding rows in the input matrix. The masked input \(\tilde{\mathbf{H}}_{t}^{\text{norm}}\) is computed as
\begin{equation}
\tilde{\mathbf{H}}_{t}^{\text{norm}} = \mathbf{H}_{t}^{\text{norm}} \odot \mathbf{m}.
\end{equation}
where \(\odot\) is element-wise multiplication. It is constructed by comparing the maximum peak magnitude of each normalized CIR (per antenna) against a threshold \(\gamma\). If the peak exceeds \(\gamma\), the antenna is considered LoS (\(m_i = 1\)); otherwise, it is suppressed as likely NLoS (\(m_i = 0\)). The threshold $\gamma$ is selected empirically based on analysis of a labeled dataset of both LoS and NLoS CIRs. As illustrated in Fig.~\ref{fig:losnlos_cir_hist}a, LoS measurements consistently exhibit stronger peaks than NLoS counterparts. Also, based on the normalized peak power histogram in Fig. \ref{fig:losnlos_cir_hist}b, a threshold of $\gamma=0.4$ proves to be reliable for distinguishing LoS from NLoS CIR in this dataset. This threshold can be adjusted as needed by analyzing the training data specific to each testbed.
Subsequently, a mapping function \(f_{\theta}(\cdot)\), realized by a multi-layer CNN encoder parameterized by \(\theta\), maps the filtered CIR input to a known 2D position label $(x_t, y_t)$:
\begin{equation}
(x_t, y_t) = \hat{f}_{\theta}(\tilde{\mathbf{H}}_{t}^{\text{norm}}).
\end{equation}
Table \ref{tab:embedding_model_summary} describes the layers of the embedding CNN on AI/ML host of Fig.\ref{fig:fp_diagram} with results discussed in Section \ref{sec:results}.
\begin{table}[t]
\centering
\captionsetup{size=small} 
\caption{CNN Embedding Model Architecture Summary}
\label{tab:embedding_model_summary}
\small 
\begin{tabularx}{\columnwidth}{@{}Xlll@{}}
\toprule
Layer & Output Dimension & Kernel Size & Activation \\ \midrule
Conv2D & (32, 16, 100) & (3, 3) & ReLU \\
Conv2D & (64, 16, 100) & (3, 3) & ReLU \\
Flatten & (1, 102400) & - & - \\
Fully Con. & (1, 512) & - & ReLU \\
Fully Con. & (1, 128) & - & ReLU \\
Fully Con. & (1, 2) & - & - \\ \bottomrule
\end{tabularx}
\end{table}
\section{Results}\label{sec:results}
\begin{figure}[t]
    \centering
    \includegraphics[width=0.99\linewidth]{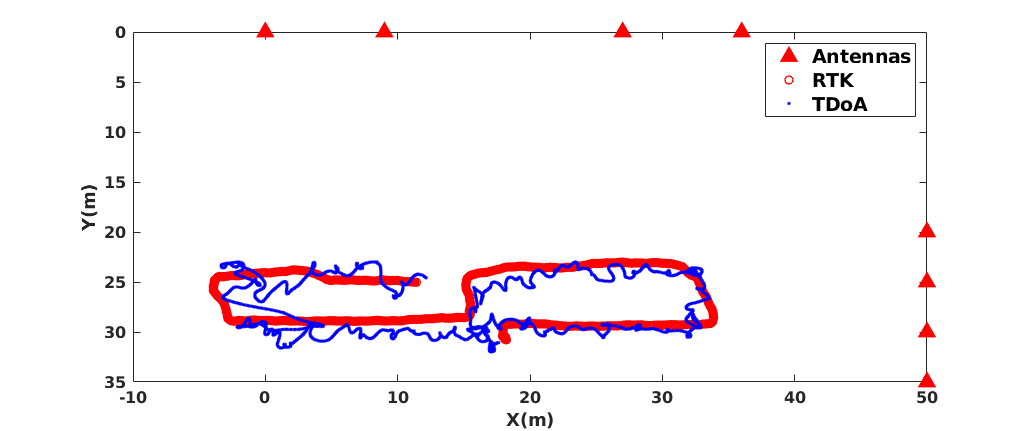}
    \caption{Benchmarking TDoA-based handheld mobile UE positioning against RTK at GEO-5G testbed}
    \label{fig:Handheld}
\end{figure}

\begin{figure}[t]
    \centering
    \begin{subfigure}[b]{0.48\linewidth}
        \centering
        \includegraphics[width=\linewidth]{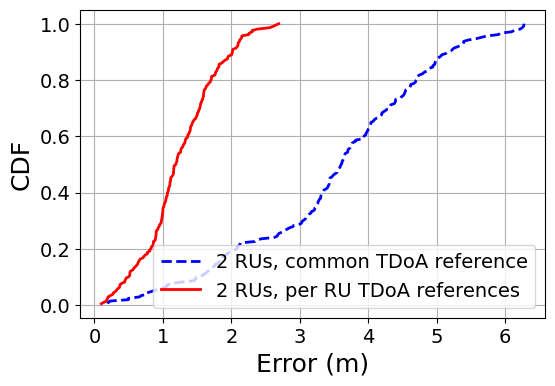}
        \caption{Common vs per-RU TDoA reference}
        \label{fig:tdoa_reference}
    \end{subfigure}
    \begin{subfigure}[b]{0.48\linewidth}
        \centering
        \includegraphics[width=\linewidth]{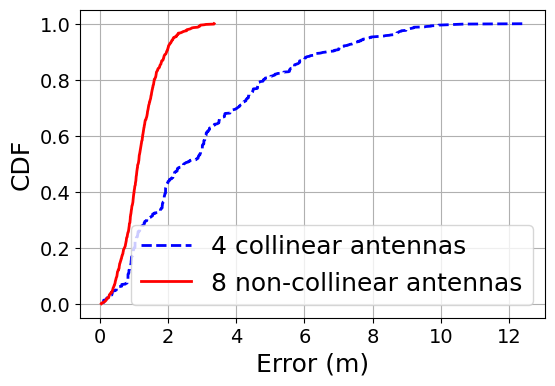}
        \caption{Collinear and non-collinear antennas}
        \label{fig:4vs8trps}
    \end{subfigure}
    \caption{Positioning performance at the EURECOM testbed}
    \label{fig:combined1}
        \Description{}
\end{figure}

This section presents experimental results from the three testbeds demonstrating the effectiveness of our proposed positioning pipeline at the LMF using TDoA and Finger Printing methods.
Starting with the GEO-5G testbed at EURECOM, 
Fig.~\ref{fig:Handheld} shows the benchmarking of our TDoA-based method against the RTK setup in a scenario where a person carries the UE along the depicted trajectory. Our method achieves 2.36m accuracy in 90\% of the time in a mobile scenario. The mobile scenario dataset is available online at \footnote{\url{https://gitlab.eurecom.fr/ahadi/5g-srs-datasets}}.
Fig.~\ref{fig:tdoa_reference} demonstrates that using per-RU TDoA references improves positioning accuracy under synchronization impairments by mitigating intra-RU timing offset. Fig.~\ref{fig:4vs8trps} compares antenna placement strategies at EURECOM, showing that 2 RUs with 8 non-collinear antennas yield more diverse TDoA measurements along both x and y-axis and better positioning accuracy than a single RU with 4 collinear antennas. 

\begin{figure}[t]
    \centering
    \begin{subfigure}[b]{0.48\linewidth}
        \centering
        \includegraphics[width=\linewidth]{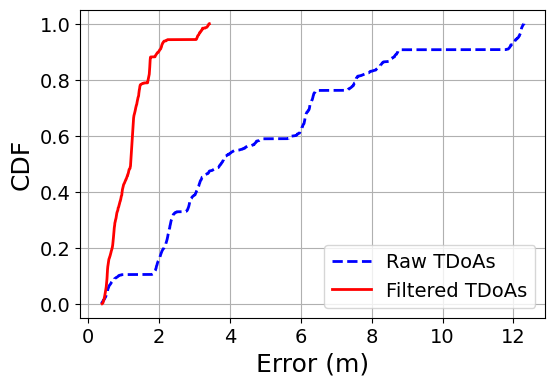}
        \caption{Filtered TDoA positioning}
        \label{fig:raw_filtered_tdoa}
    \end{subfigure}
    \begin{subfigure}[b]{0.48\linewidth}
        \centering
        \includegraphics[width=\linewidth]{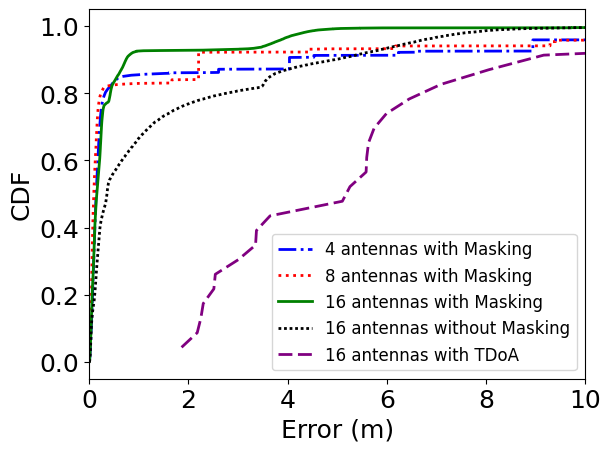}
        \caption{Finger Printing}
        \label{fig:fp_cdf}
    \end{subfigure}
    \caption{Positioning at (a)Stellantis and (b)Airbus}
    \label{fig:combined2}
        \Description{}
\end{figure}

Fig.~\ref{fig:raw_filtered_tdoa} presents results from Mattern Lab where dense multipath was observed. This figure shows that our TDoA filtering significantly reduces outlier measurements and improves overall positioning accuracy. These results support the importance of discarding unreliable TDoAs, especially under dense multipath conditions.

Fig.~\ref{fig:fp_cdf} shows the performance gain of the proposed Finger Printing method with NLoS masking, tested on a mixed LoS/NLoS dataset collected at Airbus. Compared to conventional Finger Printing, masking NLoS CIRs improves accuracy by reducing noise in the training and inference stages. The figure also demonstrates the resilience of the Finger Printing model when trained with fewer antennas, showing only a moderate drop in performance. This highlights a major advantage of AI/ML-based methods as they are less dependent on antenna count and can generalize better from large, labeled datasets. The summary of all numerical results is presented with Mean Absolute Error (MAE) and 90 percentile in error CDF (CE90) in meters in Table\ref{tab:results_comparison}.
\begin{table}[t]
\centering
\renewcommand{\arraystretch}{0.85}
\small
\begin{tabular}{|c|c|c|c|}
\hline
\textbf{Testbed} & \textbf{Conditions} & \textbf{MAE (m)} & \textbf{CE90 (m)} \\
\hline
\multirow{4}{*}{GEO-5G}     
    & 4 collinear antennas TDoA     & 3.12 & 6.67 \\ 
    & 8 non-collinear antennas TDoA & 1.20 & 1.96 \\ 
    & Common reference TDoA         & 3.12 & 5.17 \\
    & Per RU reference TDoA         & 1.20 & 2.02 \\
    & Handheld mobile UE TDoA         & 1.32 & 2.36 \\ \hline
\multirow{2}{*}{Stellantis} 
    & Filtered TDoA   & 1.22 & 1.99 \\
    & Unfiltered TDoA & 4.81 & 8.74 \\ \hline
\multirow{4}{*}{Airbus}     
    & 16 antenna TDoA & 5.78 & 8.94 \\
    & 16 antenna Finger Printing   & 0.54 & 0.74 \\
    & 8 antenna Finger Printing    & 1.14 & 2.21 \\
    & 4 antenna Finger Printing    & 1.51 & 4.04 \\ \hline
\end{tabular}
\caption{MAE and CE90 across different testbeds and positioning conditions}
\label{tab:results_comparison}
\end{table}
\vspace{0mm}
\vspace{0mm}
\vspace{0mm}
\section{Conclusion and Future Work}
This paper presents an experimental evaluation of three OAI 5G positioning testbeds using uplink TDoA and newly integrated OAI functionalities, NRPPa and LMF, across diverse real-world scenarios as well as a public dataset. Our contribution is a complete positioning pipeline at LMF, including ToA and TDoA filtering and a novel PSO-based position estimator that handles timing offsets. In mixed LoS/NLoS conditions, a masked Finger Printing model is trained on CIR data and introduced within a beyond-3GPP, O-RAN-inspired framework. Results show approximately 2m accuracy in 90\% of cases under favorable conditions, while highlighting the need for richer data like CIR in challenging environments. Future work targets integration with RIC and the E2 interface for standard O-RAN deployments.
\bibliographystyle{ACM-Reference-Format}
\bibliography{references}

\end{document}